\documentclass[prl,amsmath,amssymb,twocolumn,superscriptaddress,showpacs]{revtex4}
\usepackage{graphicx}

\newcommand{\bq}{{\bf q}}
\newcommand{\bp}{{\bf p}}
\newcommand{\br}{{\bf r}}
\newcommand{\bx}{{\bf x}}

\newcommand{\bQ}{{\bf Q}}
\newcommand{\bL}{{\bf L}}

\begin{document}
\title{Pair Density Wave in the Pseudogap State of High Temperature Superconductors}
\author{Han-Dong Chen}
\affiliation{Department of Applied Physics, Stanford University, CA 94305}
\author{Oskar Vafek}
\affiliation{Department of Physics, Stanford University, CA 94305}
\author{Ali Yazdani}
\affiliation{Department of Physics and Fredrick Seitz Materials
Research Laboratory, University of Illinois at
Urbana-Champaign,Urbana, Illinois 61801} \affiliation{Geballe
Laboratory for Advanced Materials, Stanford University, Stanford,
CA 94305}
\author{Shou-Cheng Zhang}
\affiliation{Department of Physics, Stanford University, CA 94305}
\begin{abstract}
Recent scanning tunneling microscopy (STM) experiments of
Bi$_{2}$Sr$_{2}$CaCu$_{2}$O$_{8+\delta}$ have shown evidence of
real-space organization of electronic states at low energies in
the pseudogap state\cite{Vershinin2004}. We argue based on
symmetry considerations as well as model calculations that the
experimentally observed modulations are due to a density wave of
$d$-wave Cooper-pairs without global phase coherence. We show that
STM measurements can distinguish a pair-density-wave from more
typical electronic modulations such as those due to charge density
wave ordering or scattering from an onsite periodic potential.

\end{abstract}
\pacs{74.25.Jb, 74.25.Dw, 74.72.-h}
\maketitle

One of the major challenges in condensed matter physics today is
to understand the electronic phase diagram of high temperature
cuprate superconductors. At the center of the debate is the nature
of electronic states in lightly hole-doped cuprates, which has
attracted a great deal of attention because of the observation of
a pseudogap in the density of states (DOS) in these compounds
\cite{TIMUSK1999}. To understand this phenomenon, some have
focused on the smooth evolution of the pseudogap into the
superconducting gap\cite{NORMAN1998}, the anomalous high frequency
conductivity\cite{CORSON1999}, or the Nernst
effect\cite{XU2000,WANG2001,WANG2002}. It was proposed that
superconducting pairing fluctuations persist above $T_c$ in
underdoped samples\cite{DONIACH1990,EMERY1995,GESHKENBEIN1997} and
are responsible for the strange electronic behavior in this
regime.  Others have argued that in the underdoped regime the
electronic system has a strong propensity for ordering hence
proposing the possibility for antiferromagnetic
order\cite{KAMPF1990}, combined antiferromagnetic and
superconducting fluctuations\cite{ZHANG1997}, stripe order
\cite{ZAANEN1989,EMERY1997B,VOJTA1999}, charge density
wave\cite{BARZYKIN1995,CHUBUKOV1996}, staggered flux
phase\cite{WEN1996} and d-density wave\cite{CHAKRAVARTY2001}.

Recently, mapping of the electronic states with the scanning
tunneling microscope (STM) has demonstrated a link between the
observation of real space electronic patterns and the pseudogap in
DOS of underdoped Bi$_2$Sr$_2$CaCu$_2$O$_{8+\delta}$
samples\cite{Vershinin2004}. In this experiment, STM was used to
detect a dispersionless modulation of the low energy electronic
states, showing an unusual enhancement of the intensity of
modulated patterns within the pseudogap energy scale. The
pseudogap modulations bears similarity to those observed near the
vortex cores\cite{HOFFMAN2002}, but seems to be distinct from the
STM features in the superconducting
state\cite{HOWALD2003,MCELROY2003,Vershinin2004}, which exhibit
energy dispersion. The origin of DOS modulation in the
superconducting state has been for the most part attributed to
quantum interference of quasi-particles\cite{WANG2003}, although
potential relevance of electronic ordering has also been
considered\cite{CHEN2002,HOWALD2003,Sachdev2003,PODOLSKY2003,KIVELSON2003,VOJTA2002}.
In Refs.\cite{HOWALD2003,PODOLSKY2003}, it was argued that the
energy dependence of the tunneling intensity can be used to
identify the type of ordering in the superconducting state.

In this Letter, we focus on the STM resolved modulations of the
DOS in the pseudogap state and propose that these are a
consequence of a novel density wave associated with $d$-wave
Cooper pairs, first proposed in the context of the charge ordering
in the vicinity of vortices\cite{CHEN2002,HOFFMAN2002}. This
state, illustrated in Fig. (1) of Ref. \cite{CHEN2002}, has a
rotationally symmetric charge periodicity of $4a\times 4a$ near
doping level $x=1/8$. This state also arises naturally in the
plaquette boson approach of Altman and Auerbach\cite{ALTMAN2002A}
or in the large $N$ approximation of the $t-J$
model\cite{VOJTA2002}. It differs from the stripe state in terms
of the rotation symmetry, and differs from the Wigner crystal
state of the holes, which would have a charge periodicity of
$\sqrt{8}a\times \sqrt{8}a$ at the same doping level. We argue
based on both symmetry and model calculations that a
pair-density-wave without global phase coherence provides a
natural explanation of the energy-dependence of experimentally
observed patterns in the pseudogap state\cite{Vershinin2004}.
Furthermore, we propose analysis of STM data which will allow a
pair-density-wave to be clearly distinguished from the typical
particle-hole charge density wave (CDW) or other modulation due to
electron scattering from a periodic potential. This analysis
relies on the symmetry of the Fourier transform of the tunneling
density of states, $\rho(\bQ,E)$, which we predict to be
approximately symmetric for a pair-density-wave when $E
\rightarrow -E$, while in case of CDW or potential scattering
$\rho(\bQ,E)$ is approximately anti-symmetric. Our numerical
results shows that this signature of pair-density-wave (PDW) state
survives breaking of the particle-hole symmetry and provides a
sufficiently clear-cut method to test our proposal. Further
experimental confirmation of PDW scenario would not
only establish the presence of an unusual ordering phenomenon but
it would also validate the view that the pseudogap state has
indeed short range pairing correlations.

To model the experimental findings, we study quasi-particle spectrum based on
solutions of the Bogoliubov-de Gennes (BdG) equations for a $d$-wave superconductor, while taking into account
classical phase fluctuations to model the pseudogap state\cite{FRANZ1998},
\begin{eqnarray}
  \left(
  \begin{array}{cc}
  \hat{\mathcal{H}_0}& ~\hat{\Delta}\\ ~\\
  \hat{\Delta}^*~&-\hat{\mathcal{H}}_0^*
  \end{array}
  \right)
  \left(
  \begin{array}{c}
    u_n({\bf r})\\ ~\\ v_n({\bf r})
  \end{array}
  \right)
  =E_n \left(\begin{array}{c}
    u_n({\bf r})\\ ~\\ v_n({\bf r})
  \end{array}
  \right)\label{BdG-Eq}
\end{eqnarray}
with $\hat{\mathcal{H}}_0\psi_n({\bf r}) = -t\sum_\delta \psi_n({\bf r}+\delta)-\mu(\br) \psi_n({\bf r})$ and $  \hat{\Delta}\psi_n({\bf r}) = \sum_\delta   \Delta({\bf r},\delta)   \psi_n({\bf r}+\delta)$ 
subjected to the self-consistent condition
\begin{eqnarray}
  \Delta({\bf r},\delta) &=& \frac{V(\br, \br')}{2}
   \sum_{n}\left[
  u_n({\bf r})v_n^*({\bf r}')+u_n({\bf r}')v^*_n({\bf r})
  \right]\nonumber\\
  &&\quad\times
  \tanh\left(\frac{\beta E_n}{2}\right),\label{self-consistent}
\end{eqnarray}
where $\br'=\br+\delta$ and $\delta$ denote a nearest-neighbor vector, $\psi_n(\br)$ can be $u_n(\br)$ or $v_n(\br)$.  $\beta=1/k_BT$ is the
inverse of temperature $T$.
We assume $V(\br,\br')=(V(\br)+V(\br'))/2$ and focus on the periodic pairing and chemical potentials, namely, $V({\bf r})=V_0-\sum_{Q}\Delta V({\bf Q})  \cos({\bf Q}\cdot{\bf r})$ and $  \mu({\bf r})=\mu_0-\sum_{Q}\Delta\mu({\bf Q})\cos({\bf Q}\cdot{\bf r})$.

To describe the pseudogap state above $T_c$, we consider only classical
phase fluctuations i.e. the phase field $\phi(\br)$ is independent of time.
The pairing operator takes the form
\begin{equation}
 \hat{\Delta} = \frac{1}{2}
e^{i\phi(\br)/2}
\left[\Delta(\br)d(\bp)+d(\bp)\Delta(\br) \right]
e^{i\phi(\br)/2}
\end{equation}
where the differential operators $d(\bp)=2\left[\cos(p_xa)-\cos(p_ya) \right]$.
In the superconducting state, the phase field $\phi(\br)$ is rigid while in the
pseudogap its fluctuations destroy long range phase coherence. Note that in the
superconducting state, even in the presence of pairing modulations, the spectrum
remains gapless at four points on the Fermi surface.

The local DOS (LDOS) is a functional of the phase configuration, $\phi(\bx)$, given by
\begin{eqnarray}
  \rho(\br,E, \{\phi(\bx)\})&=&-\sum_n
  \biggl[|u_n(\br)|^2 f'(E_n -E)\nonumber\\
  &&\quad\quad+|v_{n}(\br)|^2f'(E_{n}+E)\biggr],
\end{eqnarray}
where the $(u_n(\br),v_n(\br))$ and $E_n$ are respectively eigenfunctions and
eigenenergy of (\ref{BdG-Eq}) for the phase configuration $\phi(\bx)$.
$f'(E)=-\beta/4\cosh^2(\beta E)$
is the derivative of the Fermi distribution function.
The Fourier transform of $\rho(\br,E,\{\phi(\bx)\})$ is
\begin{eqnarray}
  \rho(\bq,E,\{\phi(\bx)\})=
  \frac{1}{L^2}\sum_{\br}\cos(\bq\cdot\br)\rho(\br,E,\{\phi(\bx)\})
\end{eqnarray}
with $L^2$ the number of the sites.

Before presenting the numerical results, we
analyze the symmetry of $\rho(\bq,E,\{\phi(\bx)\})$
when the average of chemical potential $\mu(\br)$ is zero.
First, we consider the effect of a CDW described by a periodic chemical
potential $\mu(\br)$ oscillating around zero average
with periodicity $\bL_\mu$. The Hamiltonian has the following
symmetry: if $(u_n(\br),v_n(\br))$ is an eigenstate with energy $E_n$ and phase
configuration $\phi(\bx)$, then
$(-1)^{x+y}(-v^{\ast}_n(\br+\bL_\mu/2),u^{\ast}_n(\br+\bL_\mu/2))$ is an eigenstate with
the same energy $E_n$ for the phase configuration $\phi'(\bx)=\phi(\bx+\bL_\mu/2)$.
We used the fact that the factor $(-1)^{x+y}$ introduces a minus sign to both the
hopping and
the pairing terms leaving the chemical potential term untouched, while the shift by
$\bL_\mu/2$ changes the sign of the chemical potential term. In addition the
standard Bogoliubov-de Gennes symmetry has been used.
For the superconducting state where
$\phi(\bx)$ is $\bx$-independent one can easily show $\rho(\bQ_\mu,E,\{\phi(\bx)\})$\cite{Podolsky-comment},
\begin{eqnarray}
  \rho(\bQ_\mu,E,\{\phi\})=-\rho(\bQ_\mu,-E,\{\phi\})  \label{mu-symm}
\end{eqnarray}
where $\bQ_\mu\cdot\bL_\mu=2\pi$. In a pseudogap state where the
phase coherence is destroyed, we must
carry out a weighted average over all of the classical phase  configurations.
We assume that to a good approximation the probability distribution $P(\{\phi(\bx)\})$ is
invariant under translation, {\it i.e.} $P(\{\phi(\br)\})=P(\{\phi(\br+\bL_\mu/2)\})$
(e.g. XY-model).
We thus conclude that the symmetry (\ref{mu-symm}) survives in the presence of the
classical phase fluctuations.

Now in contrast to CDW, we consider the case of pair-density-wave.
In the case of the periodic pairing potential with periodicity
$\bL_d$ and zero uniform chemical potential,
$(-1)^{x+y}(-v^{\ast}_n(\br), u^{\ast}_n(\br))$ is an eigenstate
with energy $E_n$, which makes the Fourier transform of the LDOS
an even function of energy
\begin{eqnarray}
    \rho(q,E,\{\phi(\bx)\}) = \rho(q,-E,\{\phi(\bx)\}),
\end{eqnarray}
where $q$ is an arbitrary wave vector.

In the presence of modulation of both chemical potential and d-wave pairing
potential, the above arguments do not hold in general. However, in the special case
of the chemical potential oscillating with $\bQ_{\mu} = (2\pi/8a,2\pi/8a)$ and
pairing potential with $\bQ_{d} = (2\pi/4a,0)$ and $(0,2\pi/4a)$, one
can show that the phase averaged LDOS satisfy
$\bar{\rho}(\bQ_d,E)=\bar{\rho}(\bQ_d,-E)$ while
$\bar{\rho}(\bQ_{\mu},E)=-\bar{\rho}(\bQ_{\mu},-E)$.
We believe that this case is relevant for the actual experiments in BSCCO.

The dichotomy of the local and non-local modulation
serves as our point of departure in analyzing perturbations
which break particle hole symmetry. If the departure from the symmetry
is not too large, the above arguments are still useful as
the odd or even character of $\rho(\bQ,E)$ is true approximately.
An immediate check is the tunneling density of states at the wavevector
corresponding to the BSCCO superstructure. We expect it to enter as a
local periodic scattering potential, making $\rho(\bQ_\mu,E)$ odd under
$E\rightarrow -E$.
On the other hand, the charge density wave of d-wave bosons
(see Fig.3 of Ref\cite{CHEN2002})
should serve as an effective periodic pairing potential that is inherently
non-local. Thus $\rho(\bQ_{d},E)$ should be even under $E\rightarrow -E$.

\begin{figure}
 \begin{tabular}{c}
   \includegraphics{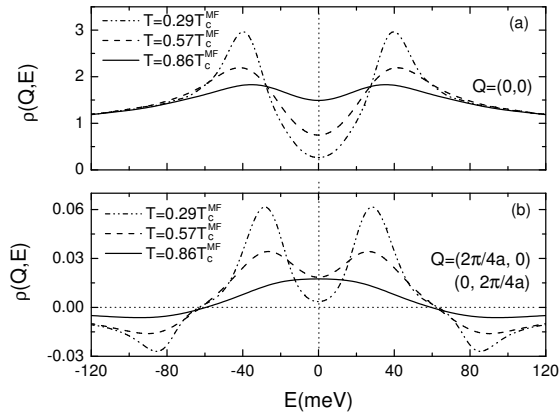}
 \end{tabular}
  \caption{(a) The average DOS. (b) The Fourier components of the LDOS at the wavevectors of the pair-density-wave.  $T^{MF}_c$ is the mean-field transition temperature.
   }\label{FIG-Pair-Modulation}
 \end{figure}

\begin{figure}
  \includegraphics{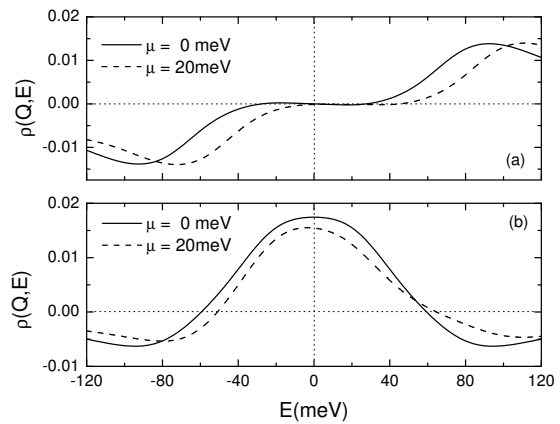}
  \caption{(a) Fourier components of LDOS with periodic chemical potential
   and uniform $d$-wave pairing. (b) Fourier components of LDOS with periodic
   $d$-wave pairing  and uniform chemical potential. The temperature is $T=0.86T_c^{MF}$ for both cases.}\label{FIG-Mu}
\end{figure}

We shall now turn to the numerical solutions of the BdG equations
(\ref{BdG-Eq}) for the superconducting state with uniform phase.
The effect of phase fluctuations will then be described later in
the text. We set $V_0=0.5t$ so which for $\mu(\br)=0$ and
$V(\br)=V_0$ yields a gap $\Delta$ of about $0.32t$ at low
temperature. We choose $t=0.125eV$ such that $\Delta=40meV$, which
is relevant for slightly underdoped BSCCO. In
Fig.\ref{FIG-Pair-Modulation} we plot the results for a
pair-density-wave with $\Delta V(\bQ_d) = 0.2 V_0$ with
$\bQ_d=(2\pi/4a,0)$ and $(0, 2\pi/4a)$. As expected,
$\rho(\bQ_d,E)$ are even functions of energy. At low temperature
compared to the mean-field transition temperature $T_c^{MF}$,
$\rho(\bQ_d,E)$ shows peaks at the energy within the
superconducting gap and zero crossings at a energies above the
gap, which is consistent with the experimental data of Howald {\it
et al.}\cite{HOWALD2003}. At higher temperature, $T=0.86T^{MF}_c$,
$\rho(\bQ_d,E)$ has a peak at zero energy and decreases as the
energy increases. Again, this is consistent with the experiment
\cite{Vershinin2004}. In the case of the periodic chemical
potential with $ \Delta\mu(\bQ_\mu)=4meV$ and
$\bQ_{\mu}=(2\pi/4a,0)$, $\rho(\bQ_{\mu},E)$ plotted in
Fig.\ref{FIG-Mu}(a) as a solid line. As expected $\rho(\bQ_\mu,E)$
is an odd function of energy. As discussed later, this feature
enables us to exclude the modulation of chemical potential as the
direct cause of the observed peak around $(2\pi/4a,0)$.

\begin{figure}[h]
  \includegraphics{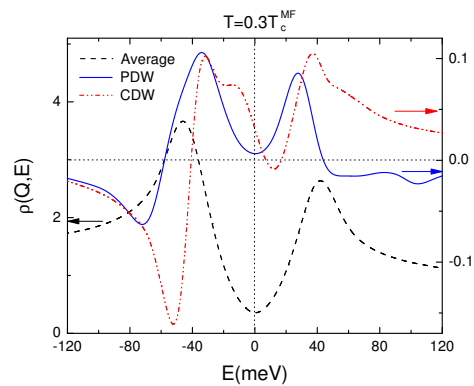}
  \caption{The Fourier components of LDOS for a more realistic model with
  next-nearest-hopping $t'=-0.3t$, chemical potential average $\mu=-1.0t$,
  the periodic modulation of pairing with wavevector $(2\pi/4a,0)$ and
  $(0, 2\pi/4a)$ and a periodic chemical potential with wavevector
 $(2\pi/10a, 2\pi/10a)$. The amplitudes of modulations are $\Delta V(\bQ_d)=0.2V_0$
  and $\Delta \mu(\bQ_\mu)=0.1t$. The dash line is the average DOS. The solid and 
  dash-dot lines are the Fourier components of the wave-vectors of pair-density-wave
  and CDW, respectively. }  \label{FIG-real-modulation-tNNN}
\end{figure}

So far, our calculation and symmetry arguments assume the particle-hole symmetry.
As an illustration of the effects of weak symmetry breaking, in Fig.\ref{FIG-Mu}
we display $\rho(\bQ,E)$ for a finite chemical potential $\mu=-20meV$.
In Fig.\ref{FIG-real-modulation-tNNN} we include finite next-nearest neighbor hopping term $t'=-0.3t$ and a chemical modulation around a finite average with
a wavevector $(2\pi/10a,2\pi/10a)$ corresponding a periodicity
about $7a$ in the $(\pi,\pi)$ direction. In both plots, remnants
of the expected symmetry are clearly visible. In particular, in
the case of the periodic $\mu(\br)$, $\rho(Q_\mu,E)$ has odd
number of zero-crossings below the gap energy, while in the case
of pair-density-wave $\rho(Q_d,E)$ has even number of
zero-crossings at energies close to the gap.

 \begin{figure}
  \includegraphics{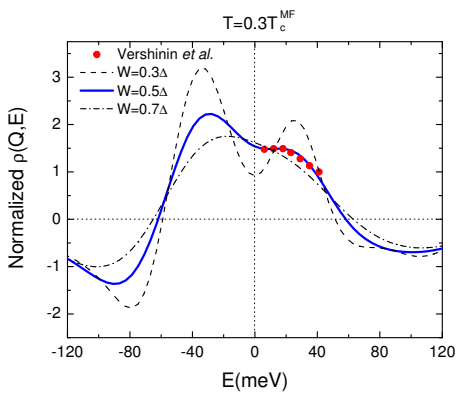}
  \caption{The LDOS in the pseudogap state calculated from Eq.(\ref{pseudogap}) and 
  Fig.\ref{FIG-real-modulation-tNNN} for $\bQ_d=(2\pi/4a,0)$ and $(0, 2\pi/4a)$.
  Here, $\Delta\approx 40meV$ is the pseudogap. The solid dots are the experimental 
  data of $(2\pi/4.7a,0)$ peak of Vershinin {\it et al}\cite{Vershinin2004}.} \label{FIG-pseudogap}
\end{figure}

 We now include the effect of the phase fluctuations within the model
 of Franz and Millis\cite{FRANZ1998},
where the LDOS of a pseudogap state $\overline{\rho}(\br,E)$ can be
calculated as an average over the phase fluctuations that are slowly varying in
space
\begin{eqnarray}
  \overline{\rho}(\br,E)=\int d\eta P(\eta)\rho(\br,E-\eta),\label{pseudogap}
\end{eqnarray}
where $\rho(\br,E)$ is the LDOS for the superconducting state, $\eta$ is just the
Doppler energy shift due to the presence of a uniform supercurrent and $P(\eta)$
is the probability distribution of $\eta$ that can be calculated within 2D $XY$
Hamiltonian to yield\cite{FRANZ1998}
\begin{eqnarray}
  P(\eta)=\sqrt{2\pi}W(T) e^{-\eta^2/2W^2(T)}
\end{eqnarray}
with $W(T)$ a temperature-dependent parameter of the order of magnitude of the
pseudogap. Since $P(\eta)$ is translationally invariant, the argument regarding
the symmetry of the LDOS is still valid.
In Fig.\ref{FIG-pseudogap}, we plot the LDOS calculated from Eq.(\ref{pseudogap}) 
 and Fig.\ref{FIG-real-modulation-tNNN} with different values of parameter $W$. 
 The curve with $W=0.5\Delta$ agrees with the experimental data of 
  Ref.\cite{Vershinin2004} reasonably well. This value of $W$
is consistent with the one obtained in Ref.\cite{FRANZ1998}.

In conclusion, we show that the Fourier transform of the
tunnelling density of states recently obtained from the STM
experiments, $\rho(Q,E)$, contains sufficient information to
distinguish the pair-density-wave state from the conventional CDW
state. Near the doping level of $x=1/8$, the pair-density-wave has
a rotationally invariant periodicity of $4a\times 4a$, which is
close to the experimental observation.
We are able to reproduce the most salient future of the non-dispersive peak as shown in Fig.2C of Vershinin {\it et al.}\cite{Vershinin2004}, namely, the intensity of the new peak picks up as the pseudogap opens. Our work also makes an
experimentally testable prediction 
about the behavior of $\rho(Q,E)$ for $E<0$, which
has the opposite behavior for the pair-density-wave state and the
conventional CDW state. The preliminary analysis of the existing data has
confirmed that the new non-dispersive peak is spatially particle-hole symmetric\cite{Misra2004}, as predicted here.
The pair-density-wave state therefore
offers a unified theoretical explanation of the STM experiments
both near the vortex core\cite{CHEN2002} and in the pseudogap
state. This state plays an essential part in understanding the
zero temperature global phase diagram of the high Tc
cuprates\cite{Chen2003}, and our current work shows that it is
also the dominant competing order in the pseudogap state. This
state could be favored in the pseudogap regime of underdoped
cuprates because of the combination of the low superfluid density
and strong pairing, which is consistent with the previous
theoretical ideas\cite{EMERY1995,Laughlin2002}.

This work is supported by the NSF under grant numbers
DMR-0342832 and the US Department of Energy, Office of Basic
Energy Sciences under contract DE-AC03-76SF00515. OV is supported
by the Stanford Institute of Theoretical Physics and HDC is
supported by a Stanford Graduate Fellowship. AY is supported under NSF (DMR-03-1529632), DOE through Fredrick Seitz Materials Research Laboratory (DEFG-02-91ER4539), ONR(N000140110071).


\end{document}